\documentclass[a4paper,12pt]{article}
\usepackage{authblk}
\usepackage{float}
\usepackage{tabularx}
\usepackage{amsmath}
\usepackage{amsfonts}
\usepackage{adjustbox}
\usepackage{multirow}
\usepackage{color}
\usepackage{longtable}
\usepackage{pdflscape}
\usepackage{rotating}
\usepackage{xcolor}
\usepackage{xcolor}
\usepackage[colorlinks = true,
            linkcolor = blue,
            urlcolor  = blue,
            citecolor = blue,
            anchorcolor = blue]{hyperref}
\usepackage[margin=1in,footskip=0.25in]{geometry}
\title{Estimating Inverse Scattering Potentials\\ for \textit{n-p} System Using \\Variational Monte Carlo \& Neural Networks}
\author{Anil Khachi$^{1}$\thanks{ anilkhachi1990@gmail.com}, Gabor Balassa$^{2*}$\thanks{ balassa.gabor@wigner.hu}
\\\vspace{0.25cm}
$^{1}$Department of Applied Science\\ Chandigarh Engineering College\\Chandigarh Group of Colleges\\ Jhanjeri, Mohali- 140307, Punjab, Bharat (India)
\vspace{0.3cm}

$^{2}$Institute for Particle and Nuclear Physics\\ Wigner Research Centre for Physics\\ Konkoly-Thege Miklós
út 29-33, 1121 Budapest, Hungary
}
\affil[*]{both authors contributed equally}
\begin{document}
\maketitle
\begin{abstract}
\noindent The Riccati-type nonlinear differential equation, also known as the Variable Phase Approach or Phase Function Method, is used to construct local inverse potentials for the \( ^3S_1 \) and \( ^1S_0 \) states of the deuteron. The Morse potential has been optimized by adjusting parameters using the Variational Monte Carlo (VMC) and Multilayer Perceptron (MLP) type Neural Networks (NN). The inverse potentials obtained from VMC and NN show almost identical parameters. In VMC, all three parameters of the Morse potential are varied to obtain the phase shifts, while in NN, the 3D-parameter optimization problem is converted to a 1D-parameter optimization problem, thus reducing optimization parameters, time, and computational cost. 

Recently, the GRANADA group published a comprehensive partial wave analysis of scattering data, which includes 6713 \( np \) phase shift data points from 1950 to 2013. Using the final experimental data points from GRANADA, we obtained the parameters for the Morse potential by minimizing the mean square error (MSE) as the cost function. The MSE using VMC (NN) is found to be 0.65 (2.5) for the \( ^1S_0 \) state and 0.16 (0.22) for the \( ^3S_1 \) state. Various quantum functions, such as phase \( \delta(r) \), amplitude \( A(r) \), and wave function \( u(r) \), are described up to 5 fm with energies \( E_{\ell ab} = [1-350 \text{ MeV}] \).
\end{abstract}

\noindent \textbf{keywords:} n-p scattering, variable phase approach (VPA), Morse potential, scattering phase shifts, Variational Monte Carlo, Neural Networks

\section{Introduction}
In the 1950s, the first paper about inverse problems emerged in fields like physics (quantum scattering theory, electrodynamics, and acoustics), geophysics (electro-, seismo-, and geomagnetic exploration), and astronomy \cite{Kabanikhin}. As computers became more powerful, these problems became useful in almost every academic field that relies on mathematical models, such as medicine, industry, ecology, economics, linguistics, and social sciences. To understand the structure of the nucleus, one of the essential methods involved is scattering. Under scattering, phase shifts play a vital role in partial wave analysis that further helps in obtaining the scattering amplitude by which the elastic scattering properties of the physical system can be calculated.

Under the ``standard approach" the main objective of theoretical physicists is to calculate the wavefunction from which other scattering properties like phase shifts, low energy scattering
parameters, static properties, etc., can be obtained. It is to be noted that the experimentalists do not get wavefunction as an output of an experiment; rather, one gets outputs like differential cross-sections and total cross-sections, transition energies, etc., which are a result of the interaction processes. Phase shifts are further related to the cross sections. The interaction is portrayed from the experimental outputs in V(r) vs r (fm) plots. 

According to Schwinger, Landau and Smorodinsky \cite{1} ``The meeting place between theory and experiment is not the phase shifts themselves but the value of the variational parameters implied by phase shifts".
  \begin{figure}[H]
    \centering
    \includegraphics[width=1\linewidth]{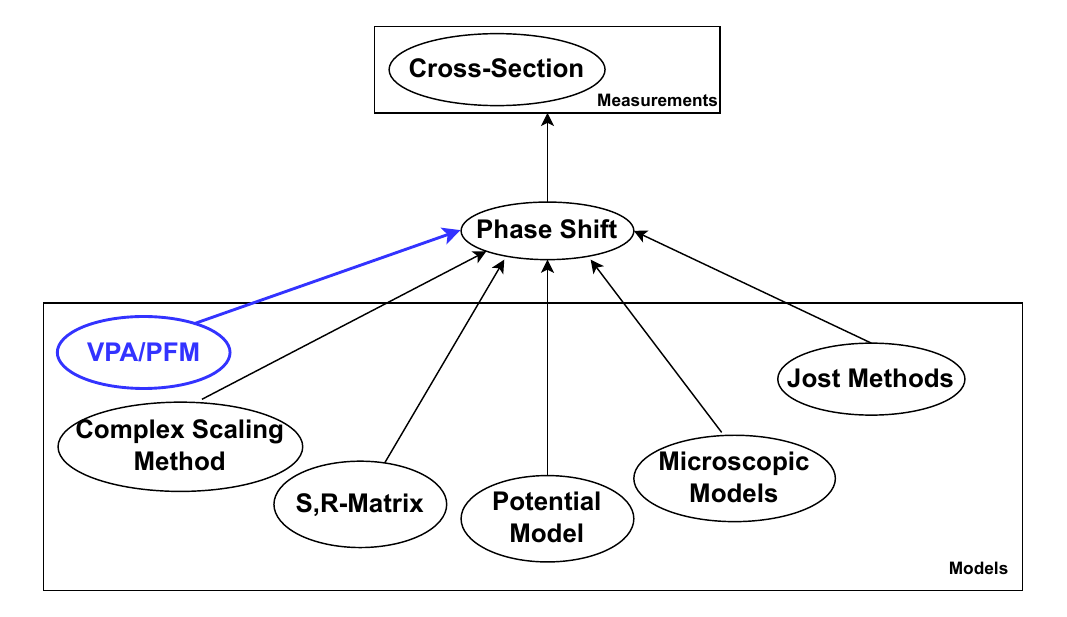}
    \caption{Different models found in the literature to obtain cross-section. In this work, we have employed a variable phase approach or phase function method (VPA/PFM) that is wavefunction independent while the rest are wavefunction dependent.}
    \label{methods}
\end{figure}

Various methods can be used to obtain the scattering phases by solving the Schr$\ddot{\text{o}}$dinger equation, like the Born approximation, Brysk's approximation, and other successive approximation techniques. These methods for phase shift and hence cross-section calculations have been shown in Fig.\ref{methods} which are wavefunction dependent. In our earlier papers, we applied VPA or PFM for studying n-p \cite{2,3,4}, n-d \cite{5} and n-$\alpha$ \cite{6} using Morse potential. In a recently published work by our group, Sastri \textit{et al.} utilized a genetic algorithm to achieve a global minimum in parameters for $\alpha-\alpha$ scattering \cite{prc-2}. Recently Blassal has applied VPA to inverse scattering problem where he used noncausal Volterra series, and multilayer perceptron (MLP) neural networks to estimate the scattering potentials for deuteron $^3S_1$ ground state \cite{balassa}. Also, Romualdi and Marchetti using VPA, presented a machine learning model based on convolutional neural networks that is capable of yielding accurate scattering phase shifts for three-dimensional spherically symmetric potentials \cite{romu}. Blassal has also applied VPA to $n-\alpha$ inverse scattering problem using neural networks \cite{Balassa}. Knowing that using VPA, phase shifts of the nucleon-nucleon scattering can be obtained using different types of potentials (high-precision phenomenological potentials) and interaction models between two nucleons. The current paper is an extension of our recently published work \cite{8} and deals with comparing the results for n-p scattering using two different optimization techniques i.e., VMC and multilayer perception type NN. Hence, the major objectives of this paper are
\begin{itemize}
\item[(i)] optimize Morse potential by two different optimization techniques i.e., variational Monte Carlo (\textit{three parameters:} $V_0$, $r_m$ \& $a_m$) and multilayer perceptron type Neural Networks (\textit{only one parameter:} only $V_0$).  
\item [(ii)] obtain various quantum mechanical functions like $\delta_0$, $A_0$ \& $u_0$ vs r for $^3S_1$ and $^1S_0$ state of deuteron
\end{itemize}
It should be noted that the true nuclear potentials are non-local, which has not been accounted for in this work and will be included later for higher partial waves of the deuteron. 
\section{Methodology}
The Morse function is given by \cite{7}
\begin{equation}
V(r) = V_0 \big(e^{-2 \frac{(r-r_m)}{a_m}}-2 e^{-\frac{(r-r_m)}{a_m}}\big)
\end{equation}
In the above equation, the parameters $V_0$, $r_m$, and $a_m$ represent the strength of interaction between the particles, the equilibrium distance at which maximum attraction is felt, and the shape of the potential, respectively. The Morse potential possesses several distinguishing properties that differentiate it from other phenomenological potentials, including:
\begin{enumerate}
    \item The Schr$\ddot{\text{o}}$dinger equation is exactly solvable for this potential.
    \item Morse function which has all the features observed in scattering phenomena such as strong repulsion at extremely short distances, an attractive nature at intermediate distances, and a quickly decaying tail for the long-range \cite{3}.
    \item Unlike other phenomenological potentials such as Manning-Rosen \cite{9}, Malfliet-Tjon \cite{10}, Hulthén \cite{11}, and others, the Morse potential has an exact analytical expression for the $^1S_0$ state phase shift.
    \item It exhibits a relatively simpler wave function \cite{matsu}.
    \item It is a shape-invariant potential. 
\end{enumerate}
Hence, it can be considered a good choice for modelling the interaction between neutron and proton.
\subsection{Variable Phase Approach (VPA) or Phase Function Method (PFM)}  
The Schr$\ddot{\text{o}}$dinger wave equation for a spinless particle with energy E and orbital angular momentum $\ell$ undergoing scattering is given by
\begin{equation}
\frac{\hbar^2}{2\mu} \bigg[\frac{d^2}{dr^2}+\big(k^2-\ell(\ell+1)/r^2\big)\bigg]u_{\ell}(k,r)=V(r)u_{\ell}(k,r)
\label{Scheq}
\end{equation}
In the above equations $k_{c.m}=\sqrt{E_{c.m}/(\hbar^2/2\mu)}$, $\hbar^2/2\mu$ = 41.47 MeV fm$^{2}$ and the energy of projectile $E_{lab}$ is converted into center of mass energy $E_{c.m.}$ for non-relativistic kinematics as $E_{c.m}=\frac{M_p}{M_n+M_p}E_{lab}$. The second-order differential equation  Eq.\ref{Scheq} has been transformed to the first-order non-homogeneous differential equation of Riccati type \cite{12,13} given by
\begin{equation}
\delta_{\ell}'(k,r)=-\frac{V(r)}{{k\left(\hbar^{2} / 2 \mu\right)}}\bigg[\cos(\delta_\ell(k,r))\hat{j}_{\ell}(kr)-\sin(\delta_\ell(k,r))\hat{\eta}_{\ell}(kr)\bigg]^2
\label{PFMeqn}
\end{equation}
 
Prime denotes differentiation of phase function with respect to distance and the Riccati Hankel function of first kind is related to $\hat{j_{\ell}}(kr)$ and $\hat{\eta_{\ell}}(kr)$ by $\hat{h}_{\ell}(r)=-\hat{\eta}_{\ell}(r)+\textit{i}~ \hat{j}_{\ell}(r)$. In integral form, the above equation can be written as
\begin{equation}
\delta(k,r)=\frac{-1}{{k\left(\hbar^{2} / 2 \mu\right)}}\int_{0}^{r}{V(r)}\bigg[\cos(\delta_{\ell}(k,r))\hat{j_{\ell}}(kr)-\sin(\delta_{\ell}(k,r))\hat{\eta_{\ell}}(kr)\bigg]^2 dr
\end{equation}

Eq.\ref{PFMeqn} is numerically solved using Runge-Kutta 5$^{th}$ order (RK-5) method \cite{14} with initial condition $\delta_{\ell}(0) = 0$. The above equation can be used to evaluate the phase function $\delta(r)$, which defines the total phase shift up to r. Then the phase shift (solution of a non-linear differential equation) is the limiting value
\begin{equation}
    \delta=\lim_{r \to \infty} \delta_\ell(r)
\end{equation}
To calculate $\delta$ the phase function $\delta(k,r)$ is evaluated up to a cut-off point at which $\delta(k,r)$ saturates or stabilizes. The physical implication of the cut-off point is that it is the radial distance beyond which the inter-nuclear potential can be considered zero. For $\ell = 0$, the Riccati-Bessel and Riccati-Neumann functions $\hat{j}_0$ and $\hat{\eta}_0$ get simplified as $\sin(kr)$ and $-\cos(kr)$, so Eq.\ref{PFMeqn}, for $\ell = 0$ becomes 
\begin{equation}
\delta'_0(k,r)=-\frac{V(r)}{{k\left(\hbar^{2} / 2 \mu\right)}}\sin^2[kr+\delta_0(k,r)]
\end{equation}
In the above equation, the function $\delta_0(k,r)$ was termed ``Phase function'' by Morse and Allis \cite{15}.
Similarly, by varying the Bessel functions for various $\ell$ values by using following recurrence relations \cite{16}
\begin{equation} 
\hat{j}_{{\ell}+1}(k r)=\frac{2 {\ell}+1}{k r} \hat{j}_{\ell}(k r)-\hat{j}_{{\ell}-1}(k r)\\ 
\end{equation}
\begin{equation}
    \hat{\eta}_{{\ell}+1}(k r)=\frac{2 {\ell}+1}{k r} \hat{\eta}_{\ell}(k r)-\hat{\eta}_{{\ell}-1}(k r)
\end{equation}
The equation for  amplitude function\cite{17} with the initial condition is obtained in the form

\begin{equation}
\begin{aligned}
    A_{\ell}^{\prime}(r) = &-\frac{A_{\ell} V(r)}{k} \left[\cos (\delta_\ell(k,r)) \hat{j}_{\ell}(kr)-\sin (\delta_\ell(k,r)) \hat{\eta}_{\ell}(kr)\right] \\
    &\times\left[\sin (\delta_\ell(k,r))( \hat{j}_{\ell}(kr)+\cos (\delta_\ell(k,r)) \hat{\eta}_{\ell}(kr)\right]
\end{aligned}
\end{equation}
for $\ell=0$ amplitude function becomes
\begin{align}
A_{0}^{\prime} &= -\frac{A_{0} V(r)}{k\left(\frac{\hbar^{2}}{2\mu}\right)} \left[\cos \delta_{0} \cdot \sin(kr) - \sin \delta_{0} \cdot (-\cos(kr))\right] \notag \\
&\quad \times \left[\sin \delta_{0} \cdot \sin(kr) + \cos \delta_{0} \cdot (-\cos(kr))\right]
\end{align}
also, the equation to obtained wavefunction\cite{17} is 
\begin{equation}
    u_{\ell}(r)=A_{\ell}(r)\left[\cos (\delta_\ell(k,r)) \hat{j}_{\ell}(k r)-\sin (\delta_\ell(k,r)) \hat{\eta}_{\ell}(k r)\right]
\end{equation}
For $\ell=0$ the wavefunction takes the following form
\begin{equation}
    u_{0}(r) = A_{0}(r) \left[\cos \delta_{0}(r) \cdot \sin(kr) - \sin \delta_{0}(r) \cdot \cos(kr)\right] 
\end{equation}

\begin{figure}[H]
    \centering
    \includegraphics[width=0.7\linewidth]{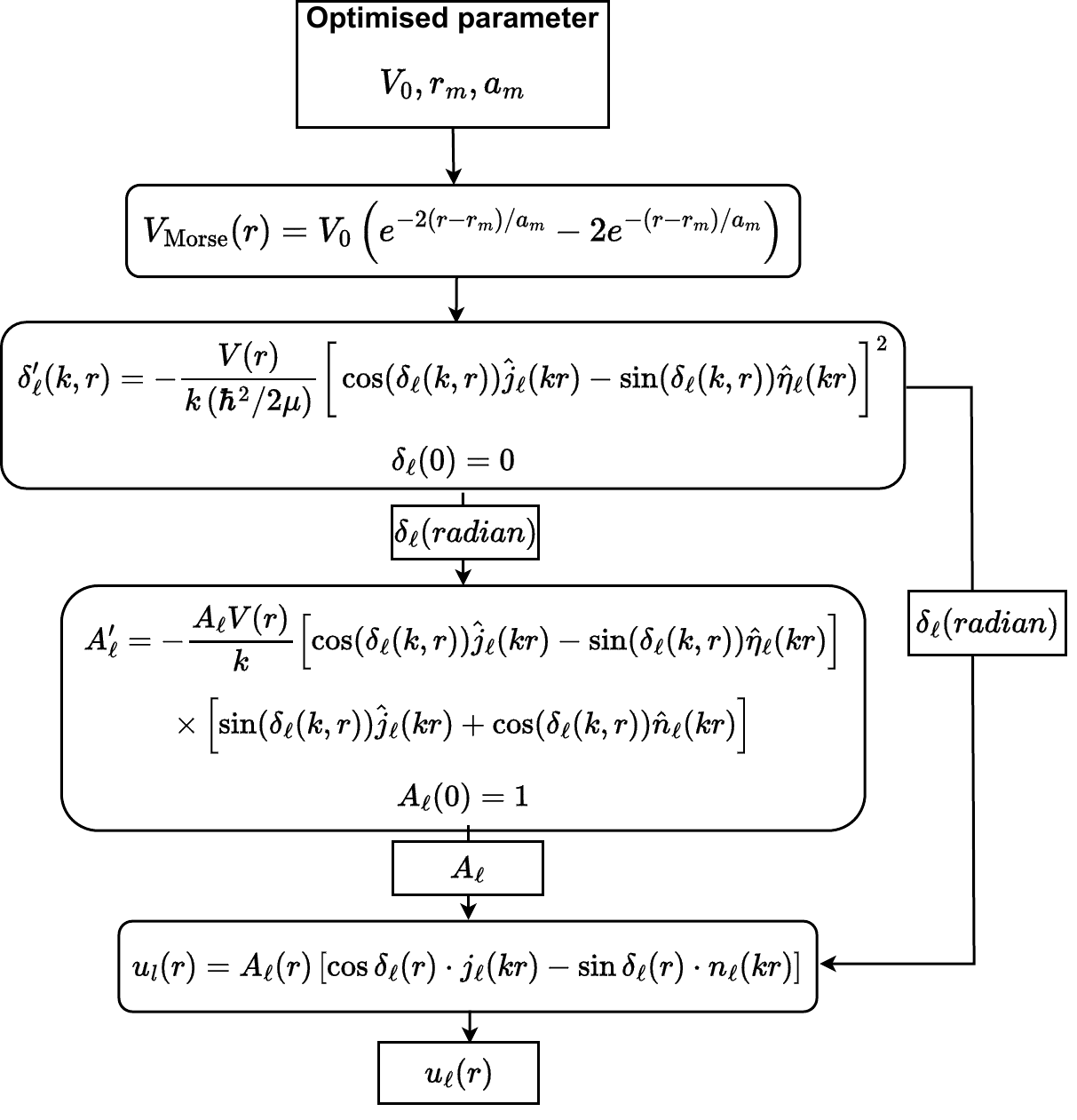}
    \caption{Detailed flowchart for obtaining phase shift, amplitude function and wavefunction for n-p system. RK5 has been used for numerical analysis.}
    \label{flowchart}
\end{figure}

Detailed flowchart to obtain inverse potential, amplitude and wavefunction has been shown in Fig. \ref{flowchart} where the first block contains model parameters, the second bock contains Morse potential, the third block contains the main VPA/PFM equation which gives $\delta$ as output, the fourth block contains amplitude function and final block contains wavefunction equation. 
\subsection{Optimization of Morse parameters using \textit{VMC}}
Typically, \textit{VMC} is utilized for obtaining the ground state energy for a given potential. The method initiates with a trial wavefunction, which is varied at a random location by a random amount in the Monte-Carlo sense. Then, the energy is determined using the newly obtained wavefunction and the variational principle is applied. This process is done iteratively till one converges to the ground state. Here, we consider varying the potential instead of the wavefunction and minimize variance with respect to experimental data, as follows:\\
\textbf{Initialisation step:} To begin the optimisation procedure, Morse parameters $V_0$, $r_m$ and $a_m$ are given some initial values. The phase equation is integrated using the RK-5 method for different values of $k$, a function of lab energies E, to obtain the simulated \textit{SPS}, say \textit{$\delta^{sim}_k$}. The mean square error (\textit{MSE}) has been determined with respect to the SPS analysis data of Wiringa \textit{et al.}, \cite{5}, say \textit{$\delta^{exp}_k$}, as\\

\begin{equation}
MSE = \frac{1}{N}\sum_{i=1}^N (\delta^{expt}_i-\delta^{sim}_i)^2
\end{equation}
This is named as $MSE_{old}$ and is also assigned to $MSE_{min}$.\\
\begin{enumerate}
\item \textbf{Initialization of model parameters:}\\
First, we create an array of scattering energies [$E_1$, $E_2$, $E_3$...] and their corresponding phase shifts [$\delta_1$, $\delta_1$, $\delta_1$...]. Then we start by choosing the initial values for the interaction potential $V_0$, i.e., $a$, $b$, $c$, etc., based on experimental data. 
\item \textbf{Solving VPA/PFM equation using RK method:}\\
In this step, we numerically solve the VPA equation (Eq. \ref{PFMeqn}) using the Runge-Kutta method (specifically, the RK-5 method used here) to obtain the simulated scattering phase shifts (SPS), which we call $\delta_{old}$. We then calculate the mean square error (MSE) between the simulated SPS and the experimental data and record it as $MSE_{old}$.
In this step, the VPA equation Eq. \ref{PFMeqn} is solved numerically using the Runge Kutta method (RK-5 method in this work) to obtain simulated scattering phase shifts (SPS) and named it as $\delta_{old}$. The mean absolute percentage error for simulated SPS w.r.t. experimental data is calculated and saved as $MSE_{old}$.
\item \textbf{Monte Carlo step:}\\
In this step, a random number, say `$r$' is generated within an interval [-I, I] and then add the perturbation `$r$' to one of the three parameters, e.g., $V_{a_{new}}$ $=V_a$ $+ r$.
\item \textbf{Recalculating the Scattering Phase Shifts:}\\
The scattering phase shift is again calculated by considering a new set of perturbed parameters, i.e., $V_{a_{new}}$. The mean absolute percentage error is again calculated and saved as $MSE_{new}$.
\item \textbf{Variational step:}\\
In this step, the condition, i.e. $MSE_{new}< MSE_{old}$ is checked. If this condition is true, then the parameter $V_a$ is updated to $V_{a_{new}}$ otherwise, the old value is retained.
\item \textbf{Iterative steps:}\\
Repeat steps 3, 4, and 5 for all the model parameters to complete one iteration. After a particular number of iterations, the size of the interval `$r$' is reduced to check if there is any further reduction in \textit{MSE}. The procedure is finished when \textit{MSE} reaches convergence, i.e. when the \textit{MSE} ceases to change.
\end{enumerate}
\subsection{Optimization Using Neural Networks}

Parameter optimization of more than two free parameters could be, in general, a very time-consuming task if we do not have any preliminary knowledge, at least on the order of magnitude of the parameters. In the case of the variable phase approximation with Morse potentials, we would need to identify three parameters that seemingly do not depend on each other in any way. In the case of the ${}^3S_1$ triplet state, we have seen that the problem could be reduced to a 2-parameter optimization problem due to the dependency of $V_0$ on $a_m$ \cite{4}. Following on this, we could assume that even if there is no analytic relation between all of the parameters in the entire range they span, there could be approximate relations between them in some well-defined region in $\mathcal{R}^3$, which we will call the operating range of the approximation. As we are dealing with the quantum scattering problem, where the available measured quantities are the asymptotic phase shifts, we will use them as 'input's for the approximations. Furthermore, we also need to decide in what order we would like to estimate the parameters. Here, we will reduce the original 3-parameter problem to a 1-parameter optimization problem by keeping $V_0$ as the free parameter and finding two functions that will approximate the $a_m$ and $r_m$ parameters from the knowledge of $V_0$ and the measured asymptotic phase shifts. Therefore, we intend to identify two functions, $F_1()$, and $F_2()$ as follows:
\begin{equation}
\label{eq:NN1}
\tilde{a}_m = F_1(V_0,\delta(k_1),\delta(k_2),...,\delta(k_n)),
\end{equation} 
and
\begin{equation}
\label{eq:NN2}
\tilde{r}_m = F_2(\tilde{a}_m,V_0,\delta(k_1),\delta(k_2),...,\delta(k_n)),
\end{equation} 
where $\delta(k_i)$ is a'measured' asymptotic phase shift at $k_i$ momentum, while $\tilde{a}_m$ and $\tilde{r}_m$ are the estimated parameters. As it can be seen, here we approach the problem in two steps, where the first step consists of the estimation of the $a_m$ parameter from the knowledge of the $V_0$ parameter, and the $\delta(k_i)$ phase shifts. The estimated parameter $\tilde{a}_m$ is assumed to be close to its true value $\tilde{a}_m \approx a_m$, however, we do not need it to be exactly the same because in a real-life experiment, we will also have a certain amount of noise in the measured phase shifts as well. Due to the possible noise in the parameters, we will also use $\tilde{a}_m$ for further estimations, where we will try to find the $F_2()$ function, which will give us an estimated value for the last parameter $r_m$. Therefore, in $F_2()$, we will use the previously approximated $\tilde{a}_m$ parameter as an input.

In the following, we have to decide on how to find the $F_1()$ and $F_2()$ functions. In general, we need a nonlinear relationship between many parameters for which neural networks could provide a great and easy way to approach the problem, therefore, we choose $F_1$ and $F_2$ to be two distinct multilayer perceptron (MLP) type neural networks, with the inputs and outputs shown in Eq.~\ref{eq:NN1} and in Eq.~\ref{eq:NN2}. To train the two MLP's, we have to generate a certain amount of data with the variable phase approximation, and then divide it for training, validation, and testing purposes. As it is impossible to cover the whole range of phase shifts and input parameters, it is necessary to set an operating range for our approximation by constraining some of the data. In this case, we have many options, e.g., one could constraint the interval of the phase shifts so that we only use potentials during the training, which will result in phase shifts in a certain range, e.g., [$0,\pi$]. By knowing the range of phase shifts we would like to use in the model, this method could greatly simplify the training procedure. Another possibility would be to simply constrain the input parameters $V_0$, $a_m$, $r_m$ of the Morse potentials. Both methods would require some preliminary knowledge of the problem to which we would like to apply the model, however, in this case, we have more than enough knowledge of the system. First of all, we know that in s-wave nucleon-nucleon scattering, the effective range of the potential is around a few fermi in distance, and not considering the possible large repulsive core at small distances, the effective magnitude of the potentials should not exceed a few hundreds of MeV's. We also know that due to the lack of many bound states, the magnitude of the phase shifts is somewhere between [$-2\pi,2\pi$]. From the phase shift measurements, which we will use in our estimations later, we could constraint the phase shifts much more, however, to keep a certain kind of 'overhead' for the network, we will try not to constraint the training too much. This will be useful in keeping good generalization properties later.

Considering all the above, the estimation of the $F_1()$, and $F_2()$ functions by MLP networks will go as follows:
\begin{itemize}
\item Generate $10.000$ Morse potentials with the parameters chosen from $V_0 \in [-250,250]$ MeV, $a_m \in [0,1]$ fm, and $r_m \in [0,5]$ fm. These parameters will give many different but still physically sensible potentials in the phase shift range we are interested in.
\item Calculate the phase shifts for the generated potentials at incident energies of $E = 0.1,1,5,10,25,50,100,150,200,250,300$ MeV. The phase shift at these values will be the input parameters for the $F_1()$, and $F_2()$ functions, and the energies correspond to the measurements.
\item Normalize the inputs and outputs between $[-1,1]$, then train the first MLP (corresponding to the $F_1()$ function) with the normalized $V_0$, $\delta(k_i)$, and $a_m$ values.
\item Using the response of the first MLP ($\tilde{a}_m$) train the second MLP, which then again corresponds to the $F_2()$ function.
\end{itemize}

\begin{figure}[!h]
\centering\includegraphics[width=5in]{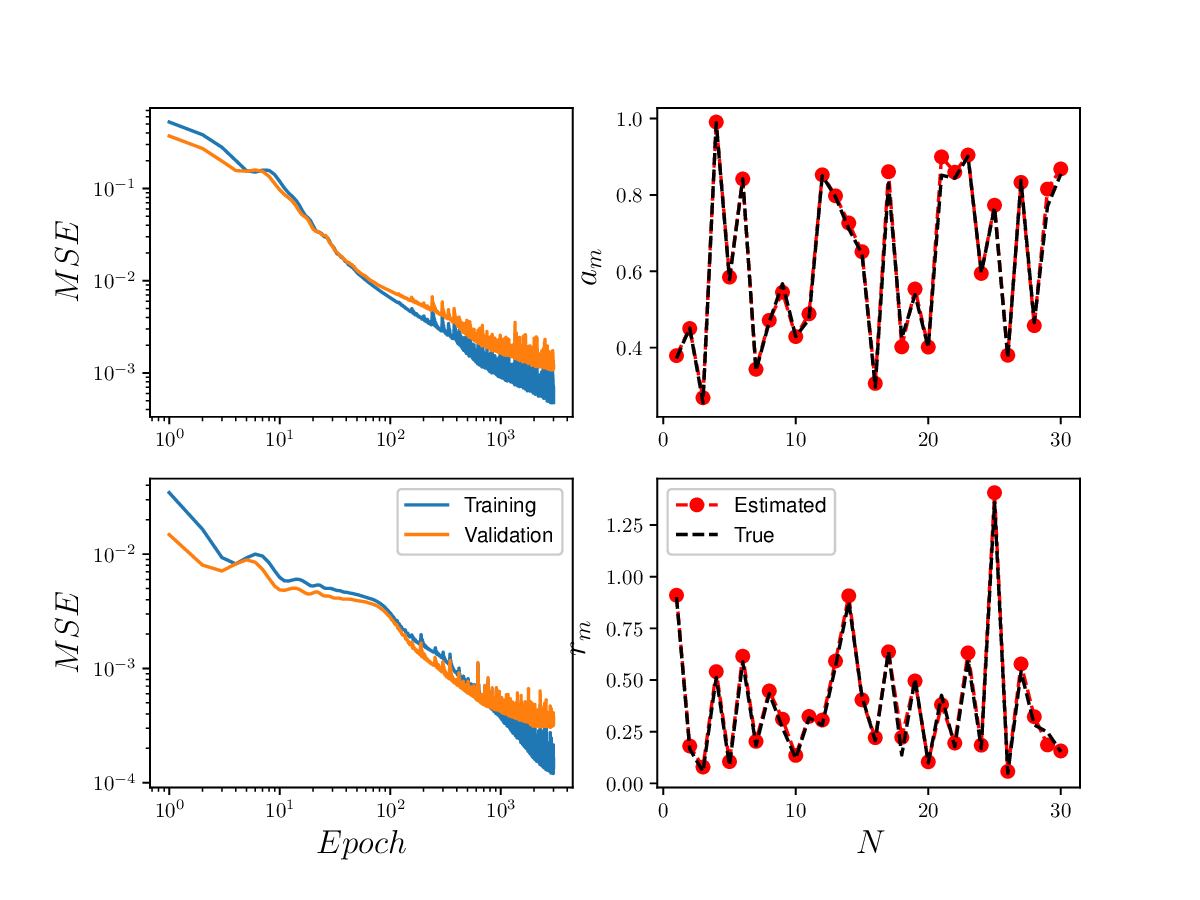}
\caption{MSE and comparison of the true and estimated values of the two neural networks corresponding to the $F_1()$ (upper row), and $F_2()$ (bottom row) functions. On the left side, the mean squared error is shown during training for the training and validation data, while on the right side, the outputs of the networks are compared to the true values for the $30$ test data (normalized). }
\label{fig:MLP_S}
\end{figure}

By training the network this way, we will be able to reduce the original $3$-parameter optimization problem to only $1$ with $V_0$ as the only free parameter, as each $V_0$ with a specific $\delta(k_1)$, $\delta(k_2)$, ..., $\delta(k_n)$ will correspond to a specific $\tilde{a}_m$, and $\tilde{r}_m$. Another important task still remains, namely, determining the complexity of the two MLP's, which in this case means the number of hidden layers and the number of neurons in each hidden layer. As there is no exact method to determine this, we rely on trying out different networks and choosing one that has some optimal behaviour for us. In our case, it was important to be able to reach good generalization relatively quickly and optimize the network like that. By trying out different configurations, we set the two MLP's so that they both have $4$ hidden layers, each having $200$ neurons with a rectified linear unit (ReLU) as the activation function. The generated 10.000 samples were divided so that we used $7000$ for training, $2000$ for validation, and the remaining $1000$ for testing. The results for the training of both networks can be seen in Fig.~\ref{fig:MLP_S}, where the upper row corresponds to $F_1()$, while the lower row corresponds to $F_2()$. On the left side, the mean squared error is shown for both the training and validation sets during the training procedure, while on the right side, the estimated values (of the normalized $a_m$, and $r_m$) are compared to the true values for the $30$ test data.

\begin{figure}[!h]
\centering\includegraphics[width=4.5in]{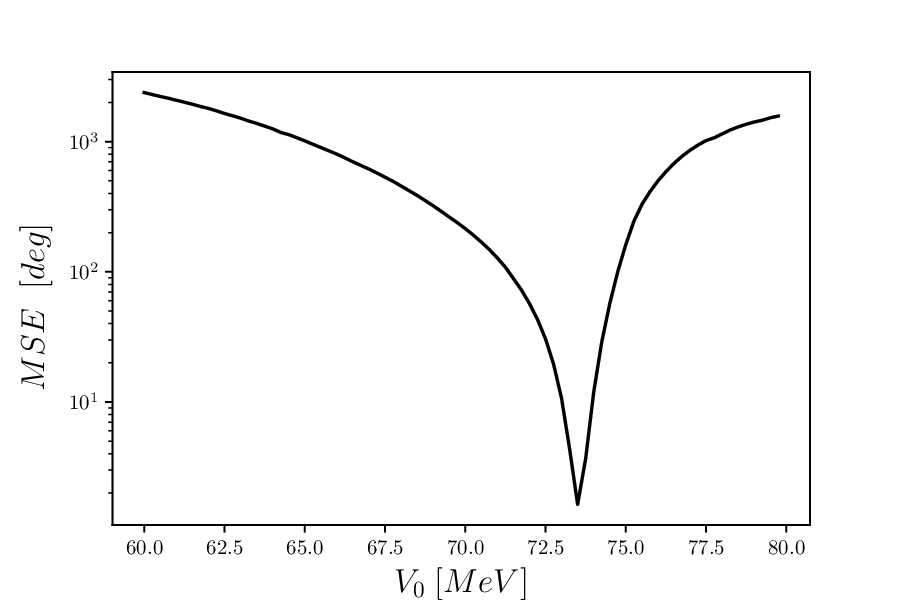}
\caption{MSE for the ${}^1S_0$ scattering scenario during the $V_0$ parameter sweep with $\Delta V_0=0.25$ MeV. A clear minimum can be seen at $V_0 = 73.5$ MeV, which corresponds to $\tilde{a}_m=0.363$ fm, and $\tilde{r}_m=0.881$ fm.}
\label{fig:1S0}       
\end{figure}

\begin{figure}[!h]
\centering\includegraphics[width=4.5in]{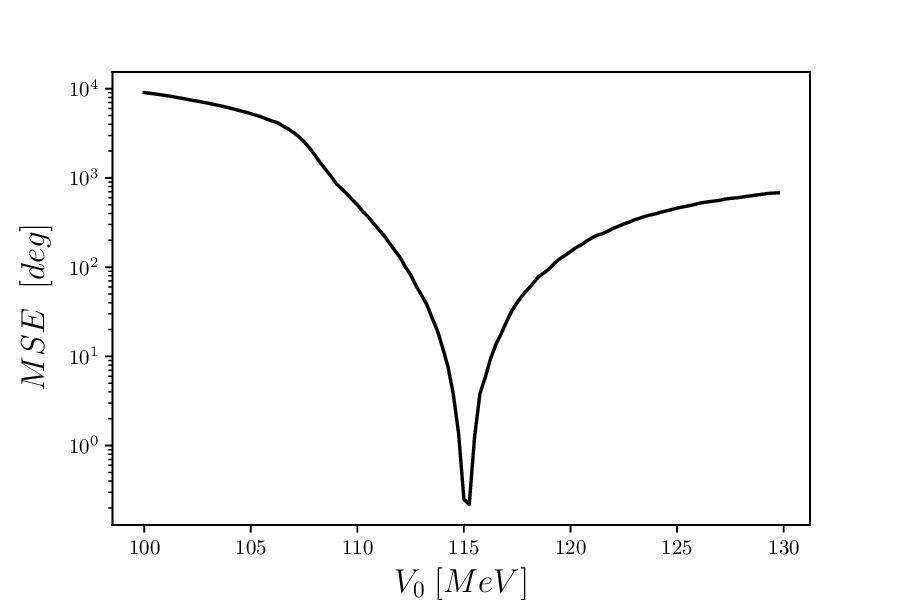}
\caption{MSE for the ${}^3S_1$ scattering scenario during the $V_0$ parameter sweep with $\Delta V_0=0.25$ MeV. A clear minimum can be seen at $V_0 = 115.25$ MeV, which corresponds to $\tilde{a}_m=0.347$ fm, and $\tilde{r}_m=0.832$ fm.}
\label{fig:3S1}       
\end{figure}

As can be seen from the test data, both networks were able to reach a very good generalization, corresponding to only a few percent relative error. By applying the trained neural networks, it is now possible to estimate the interaction potentials for s-wave scattering and compare the obtained parameters to the previously determined ones. To do this, the measured ${}^1S_0$ and ${}^3S_1$ nucleon-nucleon scattering phase shifts will be used as input parameters. Now we only need to adjust the values for $V_0$, so a simple parameter sweep will be enough for our purposes. The inversion process will thus go as follows:
\begin{itemize}
\item Choose a $V_0$, and calculate the corresponding $\tilde{a}_m$, and $\tilde{r}_m$ parameters by the trained neural networks using the measured phase shifts $\delta(k_1)$, ..., $\delta(k_n)$.
\item With $V_0$, $\tilde{a}_m$, and $\tilde{r}_m$ generate a Morse potential and calculate the phase shifts at the specified energies.
\item Compare the results to the measurements and calculate the mean squared error for the phase shifts defined as before.
\end{itemize}
By sweeping through $V_0$ we could map the whole parameter space, however, in each sweep, we have to check if the obtained $\tilde{a}_m$, and $\tilde{r}_m$ parameters are inside the operating range of the neural networks. In practice, this should not be a problem when a good generalization is obtained, but nevertheless, it is always important to check if the estimation is still in the region where we want to use the neural network. Following the steps described below, we have made sweeps with $\Delta V_0=0.25$ MeV for both the ${}^1S_0$ and ${}^3S_1$ scattering cases. The corresponding mean squared errors are shown in Fig.~\ref{fig:1S0} and in Fig.~\ref{fig:3S1}.

In both cases, there is a clear minimum in the mean squared errors that can be seen during the sweeps. The corresponding parameters calculated by the neural networks are summarized in Tab.~\ref{table1}. Let us also compare the re-calculated phase shifts with the obtained parameters with the measured phase shifts, which can be seen in Fig.~\ref{Fig.pot_phase} both for the ${}^1S_0$ and for the ${}^3S_1$ cases. The results are also compared to the ones obtained by the VMC method, giving a very good match in both cases.

The method described here can be used to describe other scattering scenarios as well, e.g., p- or d-wave scattering, or it can be used to model other types of potentials with more free parameters. Even though the trained networks have good generalization properties, they are still approximations of the true parameters. Therefore,  a possible extension of the model would be to give pre-estimated values to the parameters involved, and then, using those values as initial conditions, use some other optimization method, e.g., the previously described Monte-Carlo optimization, to obtain a better estimation of the true parameters.

\section{Results and Discussion}
Using the two different methods we obtained inverse potentials for deuteron $^1S_0$ and $^3S_1$ state. The obtained parameters are shown in Fig. \ref{table1} and are very close to each other. The original 3-D optimization problem using VMC has been reduced to 1-D optimization by using NN.

\begin{figure}[H]
    \centering
    \ \includegraphics[width=1\linewidth]{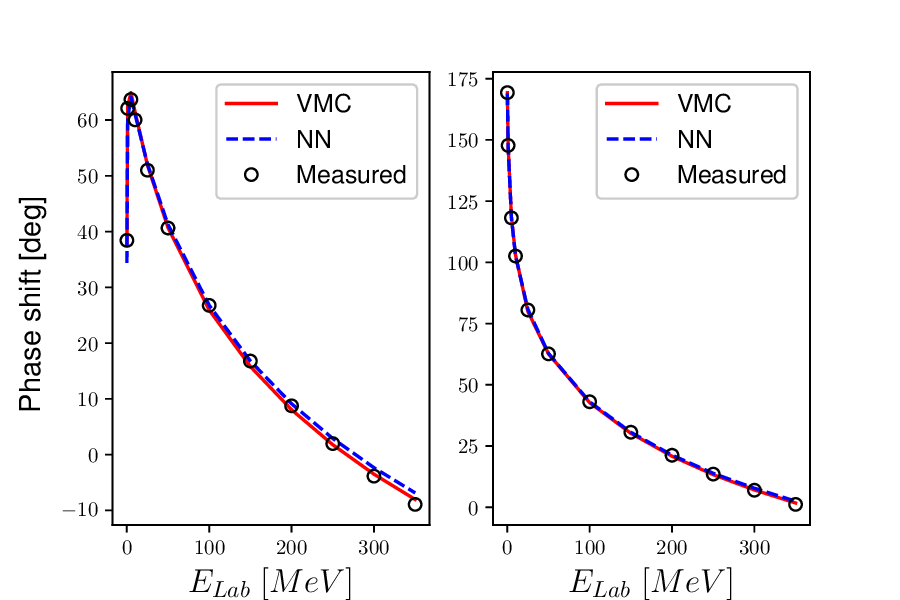}
    \ \includegraphics[width=1\linewidth]{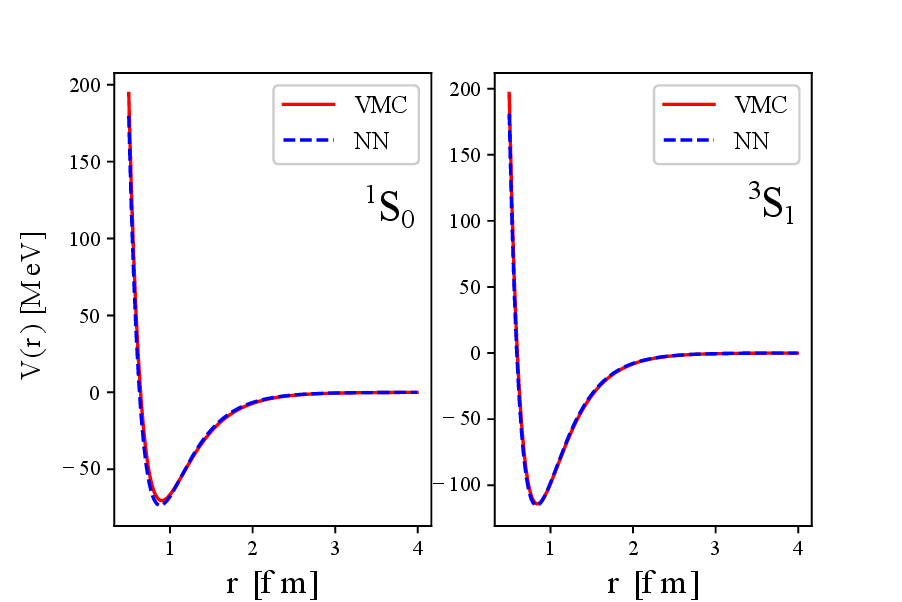}
    \caption{Top row shows scattering phase shifts for $^1S_0$ and $^3S_1$ state and bottom row shows their respective potentials obtained by the VMC, and the NN method. The optimum parameters attained are shown in table \ref{table1}.}
 \label{Fig.pot_phase}
\end{figure}

For $^1S_0$ state in the region $0 \leq r \leq 1$ \textit{fm}, the low-energy projectile faces
the attractive part of the potential. As the energy increases to $350$ MeV  the repulsive interaction takes place, indicating that the
interaction is occurring near the repulsive core. According to the VPA/PFM equation, the repulsive potential will result in negative phase shifts. This phenomenon is depicted in Fig. \ref{Fig.pot_phase} (\textit{top left}) with experimental data (black circle), estimated VMC results (solid red), and estimated neural network results (dashed blue) respectively. For $r \geq 1$ fm, the interaction moves towards the attractive core of the potential, leading to positive phase shifts. The estimated potential for $^1S_0$ state using VMC and NN is shown in Fig. \ref{Fig.pot_phase} (\textit{bottom left}) and both can be seen in very close agreement. 

\begin{table}[ht!]
\centering
\caption{$^1S_0$ \& $^3S_1$ Model Parameters using Variational Monte Carlo and Neural Networks (VMC/NN) for Morse potential. VMC parameters are taken from \cite{8}.}
\scalebox{1.0}{
\begin{tabular}{cccc|c}

\hline\hline
States      & $V_0 (MeV)$    & $r_m(fm)$  & $a_m(fm)$   & $MSE$  \\ 
\hline
 $^{1}S_{0}$ & 70.438/73.5 & 0.901/0.881 &0.372/0.363 &0.65/2.5\\ 
  $^{3}S_{1}$ & 114.153/115.25 & 0.841/0.832 &0.350/0.347 &0.16/0.22\\ 
  \hline
\end{tabular}
}
\label{table1}
\end{table}

\begin{figure}[H]
    \centering
        \includegraphics[width=0.6\linewidth]{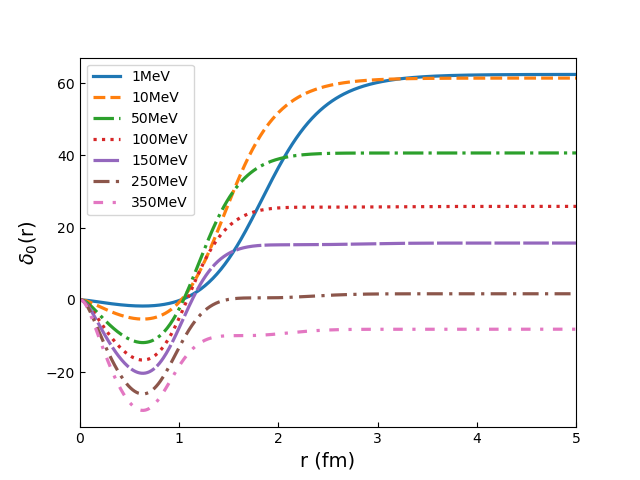}
    \centering\includegraphics[width=3.7in]{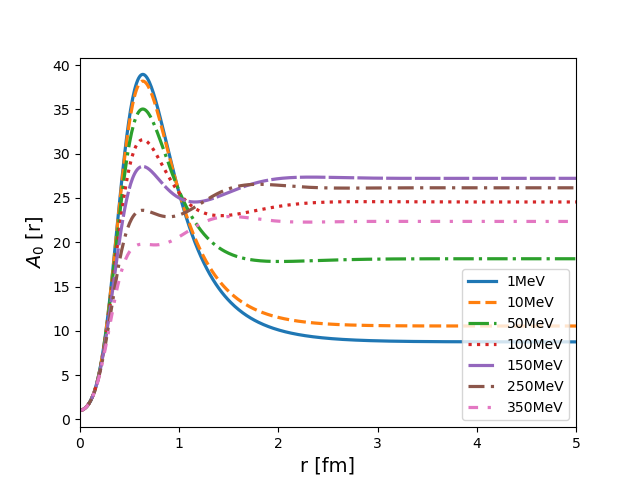}
 \includegraphics[width=0.6\linewidth]{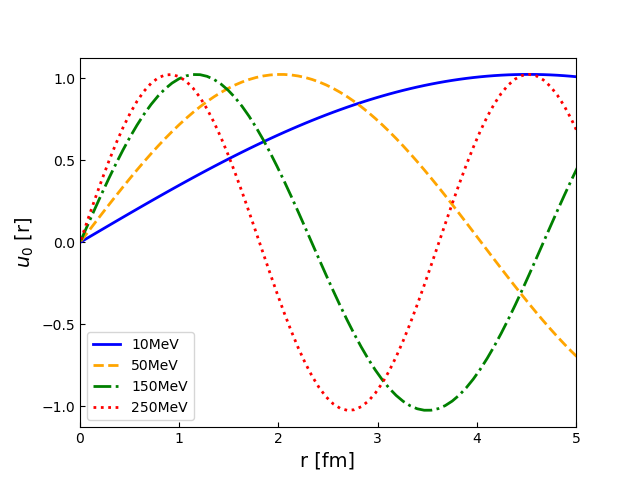}
    \caption{SPS $\delta(r)$, Amplitude $A(r)$, and wavefunction $u(r)$ for the $^1S_0$ state.}
 \label{fig4}
\end{figure}

The SPS vs r plot for $^1S_0$ and $^3S_1$ saturates after some spatial interval, depicting the vanishing of potential for larger distances. The amplitude variation for the $^1S_0$ state is shown in Fig. \ref{fig4} (\textit{mid}). Finally, the wave functions are plotted at various energies ($E = 10$, $50$, $150$, and $250$ MeV) shown in Fig. \ref{fig4} (\textit{bottom}). 

\begin{figure}[H]
    \centering
        \includegraphics[width=0.6\linewidth]{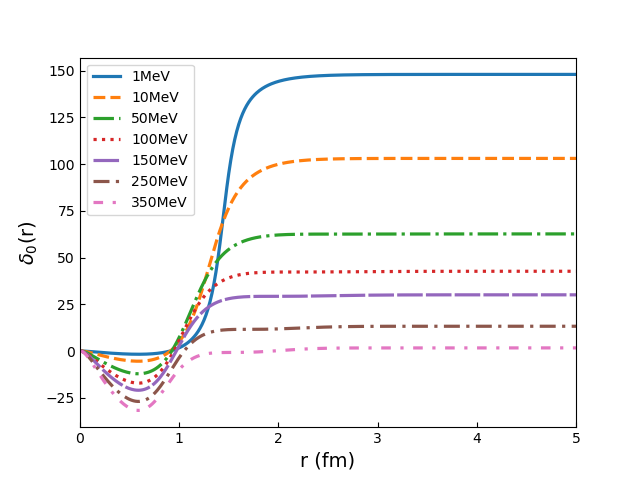}
    \includegraphics[width=0.6\linewidth]{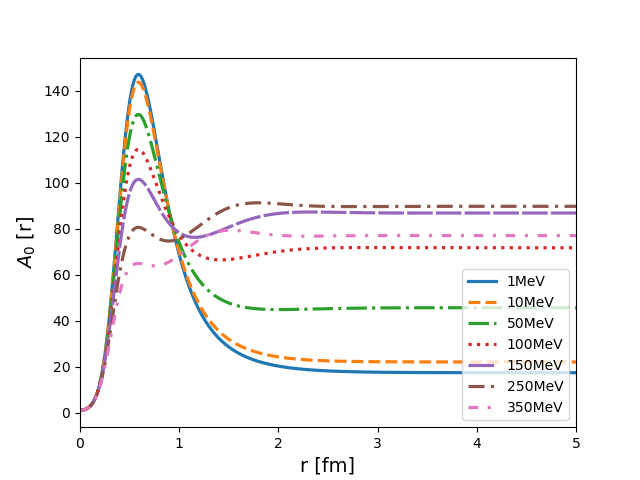}
     \includegraphics[width=0.6\linewidth]{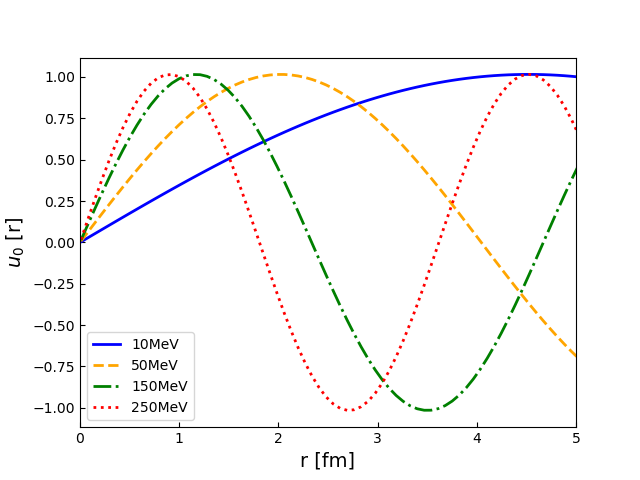}
    \caption{SPS $\delta(r)$, Amplitude $A(r)$, and wavefunction $u(r)$ for the  $^3S_1$ state.}
\label{fig6}
\end{figure}

Similarly, for $^3S_1$ state, the comparative potential using VMC and NN is shown in Fig. \ref{Fig.pot_phase} (\textit{bottom right}), SPS vs. r in Fig. \ref{fig6} (\textit{top}), amplitude function in Fig. \ref{fig6} (\textit{mid}) and wavefunction \ref{fig6} (\textit{bottom}) respectively. A similar trend like in $^1S_0$ SPS vs. r has been observed for the $^3S_1$ state, which has a much deeper potential of about 114 MeV. The jump in SPS vs r plots for the $^3S_1$ state is encountered because of the involved depth of the $^3S_1$ potential. Similarly, a more peaked amplitude is observed for the $^3S_1$ state. Model parameters obtained through VMC and NN are shown in Tab. \ref{table1}. Similarly, we have obtained the quantum mechanical functions for higher partial channels of deuteron like P and D where the optimization has been performed using VMC in tandem with VPA and the next task is to apply NN for similar work which is under process \cite{VMC}. 

Overall, the results indicate that our inverse procedure is validated by obtaining the wavefunctions in close agreement with the Av-18 potential and JISP \cite{18,19}. However, a certain phase lag in wavefunction is observed between our results and those of NIJM-II and JISP \cite{18,19}. This phase lag can be attributed to the simplistic Morse model (central potential) used in our approach. Adding realistic interactions (central \& non-central) may lead to more accurate results that are beyond the scope of the present work. Also to refine the results, we can apply the Combinatorial Data Analysis (CDA) procedure developed by our group \cite{4}. This procedure can generate a certain class of isospectral potentials that may provide a better agreement between the compared wavefunctions in this work.

\section{Conclusion}
The obtained phases serve as a quantum mechanical magnification glass, allowing us to explore the intricacies of the nuclear domain. The variation of phase shift with distance provides a beautiful demonstration that, for repulsive potentials, the phase shift $\delta$ takes on positive values, while for attractive potentials, it assumes negative values. This observation aligns perfectly with the main equation of the Variable Phase Approach. Our two different techniques, i.e., VMC and MLP type Neural Networks in tandem with VPA/PFM are effective in obtaining Phase shift $\delta(r)$, amplitude $A(r)$ and Wave function $u(r)$  for $^1S_0$ and $^3S_1$ states. Also, this work progresses in the direction of having a minimum number of parameters involved, thus saving computational time and cost.

This combined computational approach of variational Monte Carlo and Neural Networks to constructing inverse potentials offers certain advantages, although further improvements are desired for enhanced performance. Moreover, this methodology can be extended to investigate a wide range of scattering scenarios, including $n-\alpha$, $p-\alpha$, $n-C$, and higher partial waves (P, D, F, G, H, I \& J) involved in \textit{np} scattering. Also, it is well known that true nuclear interactions are known to be non-local, but that is not under the scope of the present work. The determination of a non-local potential using two different techniques for deuteron's higher partial waves will be our next objective.
\section{Acknowledgements}
A.K. is grateful to Chandigarh Group of Colleges (CGC) Jhanjeri for providing computational facilities. Also, A.K. is thankful to Dr. Ashwani Kumar Sharma (HOD Applied Physics CEC) for providing me with enough time to complete this paper.
\section*{Declaration}
The authors declare that they have no conflict of interest. 

\end{document}